\documentclass[conference]{IEEEtran}
\IEEEoverridecommandlockouts
\usepackage{lipsum}% dummy text
\usepackage{amsmath,amssymb,amsfonts}
\usepackage{algorithmic}
\usepackage{graphicx}
\usepackage{textcomp}
\usepackage{xcolor}
\usepackage{breakcites}
\usepackage{multicol, blindtext}
\usepackage{booktabs}
\usepackage{url}
\usepackage{ifthen}
\usepackage{hyperref}
\usepackage{multirow}
\usepackage{indentfirst}
\usepackage{float}
\usepackage{todonotes}
\usepackage{balance}
\usepackage{pdflscape}
\usepackage{afterpage}
\usepackage{caption}
\usepackage{subcaption}
\usepackage{array}
\usepackage{rotating}
\usepackage[utf8]{inputenc}
\usepackage[T1]{fontenc}
\usepackage{siunitx,threeparttable}
\usepackage{enumerate}
\usepackage{enumitem}
\newcommand{\PreserveBackslash}[1]{\let\temp=\\#1\let\\=\temp}
\newcolumntype{C}[1]{>{\PreserveBackslash\centering}p{#1}}
\newcolumntype{R}[1]{>{\PreserveBackslash\raggedleft}p{#1}}
\newcolumntype{L}[1]{>{\PreserveBackslash\raggedright}p{#1}}

\begin{document}

\title{Raising Secure Coding Awareness for Software Developers in the Industry}

% Choose between two different author layouts
  \author{ \IEEEauthorblockN{Tiago Gasiba}
           \IEEEauthorblockA{
              \textit{Siemens AG, München} \\
              \textit{tiago.gasiba@siemens.com}
           } \and 
           \IEEEauthorblockN{Ulrike Lechner}
           \IEEEauthorblockA{
               \textit{Universität der Bundeswehr München} \\
               \textit{ulrike.lechner@unibw.de}
           }
  }

\maketitle

\begin{abstract}

Many industrial IT security standards and policies mandate the usage of a secure coding methodology in the software development process. This implies two different aspects: first, secure coding must be based on a set of secure coding guidelines, and second software developers must be aware of these secure coding practices.
On the one side, secure coding guidelines seems a bit like a black-art: while there exist abstract guidelines that are widely accepted, low-level secure coding guidelines for different programming languages are scarce.

On the other side, once a set of secure coding guidelines is chosen, a good methodology is needed to make them known by the people which should be using them, i.e. software developers.

Motivated both by the secure coding requirements from industry standards and also by the mandate to train staff on IT security by the global industry initiative "Charter of Trust", this paper presents an overview of important research questions on how to choose secure coding guidelines and on how to raise software developer awareness for secure coding using serious games.

\end{abstract}


\section{Introduction}
\label{sec:introduction}
The Charter of Trust~\cite{SiemensCT} is a global initiative which is being undertaken by several leading companies to address the growing concerns related to IT Security of its products and services.
In order to tackle IT Security issues at its root and early stages in product development, one of the points of this initiative addresses the topic of cybersecurity education and awareness~\cite{2014_Benenson_Defining_Security_Awareness}.

This aspect is also mandated by several industry standards, to which companies are subject to compliance, such as 62.443~\cite{2018_62443_4_1}, 27k~\cite{ISO_IEC_27002:2013}, NIST~SP~800-39~\cite{NIST_SP_800_39}, etc.
As such, software developers need to be trained and familiar with how to develop, avoid security pit-falls and write secure code in the programming language being used for product development.
The basis for this is a well defined and clear set of secure coding guidelines. These come in two flavors: abstract guidelines, such as OWASP~\cite{owaspT10}, or programming language-specific such as MISRA-C~\cite{MISRA_C:2012}, CERT SEI-C~\cite{WWW_SEI_C_2018}, CERT SEI-Java~\cite{WWW_SEI_JAVA_2018}. 

In order to tackle the issue of raising IT Security awareness of software developers in the industry, our vision is to use a serious game approach where the individual challenges are based on secure coding guidelines (SCG).

This work, based on our industry experience and observations, lays out some research questions that address both the topic of selecting secure coding guidelines but also the topic on how to raise awareness about secure coding based on these guidelines.

Section~\ref{sec:state_of_the_art} outlines the current state of the art. In Section~\ref{sec:research_topics} we propose a method to derive secure coding guidelines and also present our research questions. Finally Section~\ref{sec:current_future_research} presents preliminary results and future work.
\section{State of the Art}
\label{sec:state_of_the_art}
\subsection{Secure Coding Guidelines}

Table~\ref{tab:std_reqs} shows excerpts from three prominent industry standards, which mandate secure coding practices or even explicitly the usage of secure coding guidelines. The requirement gives no clear indication about which secure coding guidelines should be adopted - this can be understood in light of the fact that there is a lacking a general consensus and standardization of SCG.

\begin{table}[H]
  \centering
  \caption{Secure Coding Requirements from Standards}
  \label{tab:std_reqs}
  \begin{tabular}{|C{1.4cm}|p{6cm}|}
    \hline
      {\bf Standard} & {\bf Requirement text} \\
    \hline
      62443-4-1 & {\it [...] incorporate security coding [...]} \\
    \hline
      27002 & {\it [...] secure  coding  guidelines  for  each  programming language used [...]} \\
    \hline
      NIST SP 800-39 & {\it [...] Information  system  security  engineers employ ... secure coding techniques [...]} \\
    \hline
  \end{tabular}
\end{table}

Our experience has shown that the quest for secure coding guidelines can result in (1) lack of SCG, (2) too many SCG or (3) conflicting SCG/recommendations.
This diversity and lack of standardization leads to companies needing to define their own set of internal accepted secure coding guidelines.
This results is a non-uniform and incoherent selection of SCG across the industry.

To the best of our knowledge, there is no previous work on how to systematically derive and define SCG (e.g. for a given programming language) and on raising awareness about SCG using serious games.
In Section~\ref{sec:research_topics} we present a proposal for a possible methodology to derive SCG.

\subsection{IT Security Awareness Training}
Software development in the industry is normally bound to a set of well established and existing programming languages~\cite{tiobe_2019}.
It has been shown that there isn't really one programming language that is significantly more secure than any another~\cite{WhiteSource2019} - vulnerabilities appear across all programming languages.

Therefore, it makes sense to focus efforts on raising awareness of software developers on how to write secure code. According to Benenson~\cite{2014_Benenson_Defining_Security_Awareness}, awareness can help to improve the understanding of the issues, to better identify the issues and to act accordingly to the issues.
Furthermore Graziotin~\cite{2018_Graziotin_Happy_Developers} has shown a correlation between developer happiness and source code quality.

One training methodology therefore that seems to be well suited is by using serious games~\cite{2016_Doerner_Serious_Games}, in particular if based on Capture-the-Flag (CTF).

\section{Research Topics}
\label{sec:research_topics}
In the previous sections, we have briefly presented the importance of secure coding guidelines both to fulfill industry standards and policies and also as a basis for IT security awareness for software developers.
Unfortunately not all programming languages have widely agreed secure coding guidelines, which leads to companies having to define their own.
In the following, we propose a method to systematically derive secure coding guidelines.
Furthermore, with the goal of raising secure coding awareness we present possible research questions to achieve this goal.

\subsection{Systematic Derivation of Secure Coding Guidelines}
Given a vulnerability database, such as~\cite{WWW_CVE}, we propose a systematic method to derive secure coding guidelines comprising the following steps:

\begin{enumerate}
    \item define a business impact metric (BIM) for vulnerabilities
    \item compute the BIM for all vulnerabilities in the database
    \item map vulnerabilities and BIM to language-specific rules
    \item compile the set of rules into secure coding guidelines
\end{enumerate}

The BIM is a company-specific metric which shall represent the perceived negative impact of the exploitation of the given vulnerability. This metric shall be aligned with business objectives and risk appetite~\cite{ISACA2016} and can include parameters such as: impact score (e.g. based on estimated money loss), probability of occurrence, perceived ease of exploitation, etc.

The mapping of vulnerabilities to language-specific rules and constructs shall be done between IT security experts and software developers.
At this stage, several language-specific recommendations could result from a single vulnerability.
The last step is a codification step, which consolidates and abstracts all the derived recommendations into a catalog of secure coding guidelines.

The main advantage of this method is that, due to the usage of a metric, the resulting secure coding guidelines can be prioritized in terms of business importance.
This leads to a natural categorization of the most important guidelines to focus on awareness training programs. 

\subsection{Secure Coding Awareness for Software Developers}
Recently, there has been an increased interest on using serious games~\cite{2016_Doerner_Serious_Games} to raise IT security awareness e.g.~\cite{Awojana2019,2017_rieb_gamified_approach,2018_rieb_IT_sicherheit}.

While the published work until now shows good indicators of the suitability of this approach, it has been (1) focused on a different target group than the one we wish to address, e.g. pentesters or security experts and (2) focused on general IT security awareness, e.g. email and password handling.

However, our target group are software developers for the industry and the content of the training is specific to secure coding.
Nevertheless, we also hypothesize that an adapted serious games of the type CTF can also be effectively used to raise secure coding awareness of software developers.
Our assumption is based on the positive indicators from similar work, but also on the following facts: (1) participants typically enjoy playing CTF games (Kees et al.~\cite{2017_kees}) and (2) happy developers write better code (Graziotin et al.~\cite{2018_Graziotin_Happy_Developers}).

\subsection{Research Questions}
\label{sub:sec:rq}
This short paper has briefly shown how important secure coding guidelines are for the industry and also for raising software developer awareness on the topic of secure coding. However, it does also raise some further important questions that need additional research. These questions include:

\begin{enumerate}
  \item[Q1]{What is the current state of usage of SCG across the industry?}
  \item[Q2]{How to can SCG be systematically derived?}
  \item[Q3]{How to raise awareness about SCG for software developers in the industry by means of CTF serious games?}
\end{enumerate}

The first research question {\it Q1}, should allow us to validate the assumption that our reported experience is also shared among the industry. 
Question {\it Q2} would help in {\it Q3} when secure coding guidelines are missing as input to create a serious game. Due to the derivation of a business metric, it also allows to rank guidelines by importance to business.
Motivated by the industry problem exemplified in this work, {\it Q3} tries to address it by means of designing a serious game.

\section{Preliminary Results and Future Research}
\label{sec:current_future_research}
Currently ongoing investigations, based on a requirements engineering approach, intend to address the questions presented in Section~\ref{sub:sec:rq}.
The result aims at contributing on how to improve IT security awareness, in particular on secure coding topics, of software developers in the industry and, as a consequence, lead to improved quality of products and services.

Preliminary results~\cite{gasiba_re19} on the requirements for Capture-the-Flag challenge design give a positive indication that defensive-style game are appropriate for raising awareness about secure coding.
Furthermore it confirms the happiness and satisfaction of the participants playing the game.
Further preliminary research suggests that the presented methodology to derive secure coding guidelines can indeed be used as input to design defensive challenges and also to plan and prioritize a teaching curriculum.

Investigations which shall address the research questions above and also the architecture of the Capture-the-Flag serious game and player engagement are currently underway.

\bibliographystyle{IEEEtran}
\bibliography{paperDB.bib}

\end{document}